\documentstyle[epsf,12pt]{article}

\begin{document}

\title{One-loop renormalization of the Yang-Mills theory with BRST-invariant
mass term.}

\author{E.Andriyash \thanks{E-mail:$anzhenya@yandex.ru$},
K.Stepanyantz \thanks{E-mail:$stepan@theor.phys.msu.su$}}

\maketitle

\begin{center}

{\em Moscow State University, physical faculty,\\
department of theoretical physics.\\$117234$, Moscow, Russia}

\end{center}

\begin{abstract}
Divergent part of the one-loop effective action for the Yang-Mills theory
in a special gauge containing forth degrees of ghost fields and allowing
addition of BRST-invariant mass term is calculated by the generalized
t'Hooft-Veltman technique. The result is BRST-invariant and defines
running mass, coupling constant and parameter of the gauge.
\end{abstract}

\sloppy

%%%%%%%%%%%%%%%%%%%%%%%%%%%%%%%%%%%%%%%%%%%%%%%%%%%%%%%%%%%%%%%%%%%%%%%%%%

\section{Introduction.}
\hspace{\parindent}

It is well known, that adding of the mass term to the action of the
Yang-Mills theory breaks the gauge invariance. That is why the massive
gauge fields are usually introduced by the spontaneous symmetry breaking
mechanism \cite{Huang}. Nevertheless, there is an alternative approach.
It is based on the fact, that the quantization of Yang-Mills theory
requires to fix the gauge invariance and add to the action ghost Lagrangian
\cite{Slavnov_Book}. The total Lagrangian is then invariant under BRST
transformations, which remain from the original gauge transformations.
In principle, it is possible to construct mass term invariant only under
these residual transformations. For a general gauge this problem is not
solved. However, there is a special gauge, containing forth degree of
ghost fields (in principle, such gauges are used sometimes in the quantum
field theory, see e.f. \cite{Kastening}), allowing existence of the
BRST-invariant mass term \cite{Curci1,Curci2,Kondo1}. The corresponding
model is not unitary \cite{Curci2,Ojima,Boer}, but was proven to be
multiplicably renormalizable \cite{Curci1,Boer,Delduc,Blasi}.

It is interesting to investigate quantum properties of such theory
and, in particular, to check BRST invariance of the effective action.
In principle, the one-loop \cite{Kondo2} and two-loop corrections
\cite{Gracey} were already found by the diagram technique. However, the
calculations are rather involved and it is desirable to check them by
another method of calculation. One of the suitable calculational tools
is t'Hooft-Veltman technique \cite{tHooft}, which was originally
proposed for a limited class of theories and afterwards generalized for
the general case in \cite{NuclPhys}. In the frames of t'Hooft-Veltman
method the divergent part of the one-loop effective action can be
calculated by some purely algebraic operations. In this paper such
calculation is made for the Yang-Mills theory with the BRST-invariant
mass term.

The paper is organized as follows: Some information about Yang-Mills
theory with BRST-invariant mass term is reminded in Section
\ref{Section_YM}. The generalized t'Hooft-Veltman technique and the
calculation process is briefly described in Section
\ref{Section_Calculations}. Renormalization of the model is discussed
in Section \ref{Section_Renormalization}. In this section we also check,
that the divergent part of the one-loop effective action is invariant
under BRST transformations. The obtained results are briefly discussed
in Conclusion and some technical details of calculations are presented
in the Appendix.

%%%%%%%%%%%%%%%%%%%%%%%%%%%%%%%%%%%%%%%%%%%%%%%%%%%%%%%%%%%%%%%%%%%%%%%%%%%

\section{Yang-Mills theory with BRST-invariant mass term.}
\hspace{\parindent}
\label{Section_YM}

If the gauge invariance in the Yang-Mills theory is fixed by adding
of the following terms \cite{Kondo1}

\begin{equation}
S_{gf} + S_{gh} = \frac{1}{e^2} \mbox{tr} \int d^4 x \Big(
\alpha B^2 - 2 \partial_\mu B\, A^\mu
- i \alpha B \{ c \, \bar c \}
- 2i \partial_\mu {\bar c} {\cal D}^\mu c + \alpha c^2 {\bar c}^2
\Big),
\end{equation}

\noindent
then it is possible to add a mass term, invariant under BRST
transformations up to a total derivative. The action of the obtained
theory is written as

\begin{eqnarray}\label{YM_Action}
&& S = \frac{1}{e^2} \mbox{tr} \int d^4 x \Big(
\frac{1}{2} F_{\mu\nu}^2 + \alpha B^2 - 2 \partial_\mu B\, A^\mu
- i \alpha B \{ c \, \bar c \}
-\nonumber\\
&&\qquad\qquad\qquad\qquad\qquad
- 2i \partial_\mu {\bar c} {\cal D}^\mu c + \alpha c^2 {\bar c}^2 -
m^2 ( A_\mu^2 - 2i \alpha c {\bar c} )   \Big).\qquad
\end{eqnarray}

\noindent
and is invariant under the following BRST transformations:

\begin{eqnarray}\label{BRST_Transformations}
&& \delta_b A_\mu(x) = \varepsilon_b {\cal D}_\mu c(x);\nonumber\\
&& \delta_b c(x) = - \varepsilon_b c(x)^2;\nonumber\\
&& \delta_b \bar c(x) = i \varepsilon_b B(x);\nonumber\\
&& \delta_b B(x) = 0.
\end{eqnarray}

\noindent
There is also an invariance under anti-BRST transformations

\begin{eqnarray}\label{BRST_Anti_BRST_Transformations}
&& \delta_a A_\mu(x) = \varepsilon_a {\cal D}_\mu \bar c(x);\nonumber\\
&& \delta_a c(x) = - \varepsilon_a (i B(x) + \{c(x), \bar c(x)\});
\nonumber\\
&& \delta_a \bar c(x) = - \varepsilon_a \bar c(x)^2;\nonumber\\
&& \delta_a B(x) = 0.
\end{eqnarray}

%%%%%%%%%%%%%%%%%%%%%%%%%%%%%%%%%%%%%%%%%%%%%%%%%%%%%%%%%%%%%%%%%%%%%%%%%%%

\section{Calculation of the divergent part of the one-loop effective
action.}
\hspace{\parindent}
\label{Section_Calculations}

For calculation of divergent part of the effective action it is
possible to use different methods, for example, diagram technique
or t'Hooft-Veltman method \cite{tHooft,NuclPhys}. In this paper we
will use the second method. Let us remind its main ideas:

It is well known \cite{Huang}, that for a theory described by a classical
action $S(\varphi_i)$ the one-loop effective action can be written as

\begin{equation}\label{Calculations_Gamma_Ln}
\Gamma^{(1)} = \frac{i}{2} \mbox{Str}\,\ln\, D_i{}^j
\end{equation}

\noindent
where

\begin{equation}
D_i{}^j = \frac{\delta^2 S}{\delta \varphi^{i}\delta \varphi_j}
\end{equation}

\noindent
is an operator of second variation of the classical action. In the most
general case this operator can be written as

\begin{eqnarray}\label{Calculations_General_Operator}
&&
\vphantom{\frac{1}{2}}
D_{i}{}^{j} =
K^{\mu_1\mu_2\ldots \mu_{L}}{}_{i}{}^{j}
\ \nabla_{\mu_1} \nabla_{\mu_2}\ldots
\nabla_{\mu_{L}}
+\ S^{\mu_1\mu_2\ldots \mu_{L-1}}{}_{i}{}^{j}
\ \nabla_{\mu_1} \nabla_{\mu_2}\ldots
\nabla_{\mu_{L-1}}
\nonumber\\
&&
\vphantom{\frac{1}{2}}
+\ W^{\mu_1\mu_2\ldots \mu_{L-2}}{}_{i}{}^{j}
\ \nabla_{\mu_1} \nabla_{\mu_2}\ldots
\nabla_{\mu_{L-2}}
+\ N^{\mu_1\mu_2\ldots \mu_{L-3}}{}_{i}{}^{j}
\ \nabla_{\mu_1} \nabla_{\mu_2}\ldots
\nabla_{\mu_{L-3}}
\nonumber\\
&&
\vphantom{\frac{1}{2}}
+\ M^{\mu_1\mu_2\ldots \mu_{L-4}}{}_{i}{}^{j}
\ \nabla_{\mu_1} \nabla_{\mu_2}\ldots
\nabla_{\mu_{L-4}} +
\ldots,
\end{eqnarray}

\noindent
where all tensors are considered to be symmetrical with respect to
permutations of Lorentz indexes and $\nabla_\mu$ denotes a covariant
derivative

\begin{eqnarray}
&&\nabla_\alpha T^\beta{}_i{}^j = \partial_\alpha T^\beta{}_i{}^j +
\Gamma_{\alpha\gamma}^\beta T^\gamma{}_i{}^j + \omega_{\alpha i}{}^k
T^\beta{}_k{}^j - T^\beta{}_i{}^k \omega_{\alpha k}{}^j;\nonumber\\
&&\vphantom{\frac{1}{2}}
\nabla_\mu \Phi_i = \partial_\mu \Phi_i + \omega_{\mu i}{}^j \Phi_j,
\end{eqnarray}

\noindent
$\Gamma_{\mu\nu}^\alpha$ is a Cristofel symbol

\begin{equation}
\Gamma_{\mu\nu}^\alpha = \frac{1}{2} g^{\alpha\beta} (\partial_\mu
g_{\nu\beta} + \partial_\nu g_{\mu\beta} - \partial_\beta g_{\mu\nu})
\end{equation}

\noindent
and $\omega_{\mu i}{}^j$ is a connection in the principle bundle.

The divergent part of one-loop effective action
(\ref{Calculations_Gamma_Ln}) for operator
(\ref{Calculations_General_Operator}) was found explicitly in
\cite{NuclPhys}. In the particular case, when operator
(\ref{Calculations_General_Operator}) has second order in derivatives
($L=2$), and the space-time is flat, the result is written as

\begin{eqnarray}\label{Calculations_Algorithm}
&& \Gamma^{(1)}_\infty  = - \frac{1}{16\pi^2} \ln\frac{M}{\mu}
\,\mbox{Str} \int d^4x\, \Big\langle
\frac{1}{2} \hat W^2
- \hat W\,\hat S^2
- 2 \partial_\mu \hat S\,\hat W\,\hat K ^\mu
+\nonumber\\
&&
+\frac{1}{4} \hat S^4
+\frac{1}{3} \Bigg(
\partial_\mu \hat S^\mu \hat S^2
- 2 \partial_\mu \hat S\,\hat K^\mu \hat S^2
- \partial_\mu \hat S\,\hat S\,\hat S^\mu
+ 2 \partial_\mu \hat S\,\hat S^2 \hat K ^\mu
\Bigg)
+\nonumber\\
&&
+ \partial_\mu \hat S\,\partial_\nu \hat S ^\nu \hat K ^\mu
- \frac{1}{2} \partial_\mu \hat S\,\partial_\nu \hat S
 \left( - \hat K ^{\mu\nu}
+ 2 \hat K^\mu \hat K ^\nu
+ 2 \hat K^\nu \hat K ^\mu \right)\Big\rangle,
\end{eqnarray}

\noindent
where the notations can be explained by the following equations:

\begin{eqnarray}
&& \hat S_i{}^j \equiv (Kn)^{-1}{}_i{}^k (Sn)_k{}^j;
\qquad\qquad\qquad\qquad\ \
\hat K^\mu{}_i{}^j \equiv (Kn)^{-1}{}_i{}^k (Kn)^\mu{}_k{}^j;
\nonumber\\
&& (Sn)_i{}^j \equiv S^{\mu_1\mu_2\ldots \mu_{L-1}}{}_{i}{}^{j}
n_{\mu_1} n_{\mu_2}\ldots n_{\mu_{L-1}};
\qquad
(Kn)^\mu{}_i{}^j \equiv K^{\mu\mu_2\ldots \mu_{L}}{}_{i}{}^{j}
n_{\mu_2}\ldots n_{\mu_L};\qquad\nonumber\\
&&\qquad\qquad\qquad\qquad\qquad\qquad
(Kn)_i{}^k (Kn)^{-1}{}_k{}^j = \delta_i^j.
\end{eqnarray}

\noindent
Here $n_\mu$ is a unit vector and the angle brackets denotes the
following operation:

\begin{eqnarray}\label{Calculations_Angle_Integration}
&&\langle n_{\mu_{1}} n_{\mu_{2}} \ldots n_{\mu_{2m}}\rangle \equiv
\frac{1}{2^{m} (m+1)!}
\nonumber\\
&&\qquad\qquad \times
\Big(g_{\mu_{1}\mu_{2}}\,g_{\mu_{3}\mu_{4}} \ldots g_{\mu_{2m-1}\mu_{2m}}
+ \mbox{permutations}\ (\mu_1 \ldots \mu_{2m})
\Big).\qquad
\end{eqnarray}

Second variation of action (\ref{YM_Action}), which defines matrixes
$K$, $S$ and $W$, is presented in the Appendix. The matrixes constructed
>from it were substituted to equation (\ref{Calculations_Algorithm}),
which gives the divergent part of the one-loop effective action. After
this operation we obtained the following result:

\begin{eqnarray}\label{Calculations_Counterterms}
&& \Gamma^{(1)}_\infty = \frac{c_2}{8\pi^2}\,\ln\frac{M}{\mu}\,
 \mbox{tr} \int d^4 x \Bigg(
\frac{1}{2} \Big(-\frac{\alpha}{2}-\frac{13}{6}\Big)
\Big(\partial_\mu A_\nu^R - \partial_\nu A_\mu^R\Big)^2
+ \Big(-\frac{3\alpha}{4} - \frac{17}{12}\Big)
\times\nonumber\\
&& \times
\Big(\partial_\mu A_\nu^R - \partial_\nu A_\mu^R\Big)
\Big(A_\mu^R A_\nu^R - A_\nu^R A_\mu^R\Big)
+ \frac{1}{2} \Big(-\alpha - \frac{2}{3}\Big)
\Big(A_\mu^R A_\nu^R - A_\nu^R A_\mu^R\Big)^2
-\nonumber\\
&& - \frac{\alpha^2}{4} (B^R)^{2}
+ \frac{\alpha}{2} \partial_\mu B^R\, A_\mu^R
+ i \frac{\alpha^2}{2} B^R \{c^R \,\bar c^R\}
+ \frac{\alpha-3}{4} m^2 (A_\mu^R)^2
- i \frac{\alpha^2}{2} m^2 c^R\, \bar c^R
-\nonumber\\
&& - \frac{3}{4} \alpha^2 (c^R)^2 (\bar c^R)^2
+ i \frac{\alpha + 3}{2} \partial_\mu \bar c^R \,\partial_\mu c^R
+ i \alpha\, \partial_\mu \bar c^R\,[A_\mu^R, c^R] \Bigg)
\end{eqnarray}

%%%%%%%%%%%%%%%%%%%%%%%%%%%%%%%%%%%%%%%%%%%%%%%%%%%%%%%%%%%%%%%%%%%%%%%%%%%

\section{One-loop renormalization ans BRST-invariance of the effective
action.}
\hspace{\parindent}
\label{Section_Renormalization}

After adding counterterms, corresponding to equation
(\ref{Calculations_Counterterms}) to the classical action
(\ref{YM_Action}), the renormalized action can be written as

\begin{eqnarray}\label{Renormalization_Renormalized_Action}
&& S_{ren} = S + \Delta S = \frac{1}{e_0^2} \mbox{tr} \int d^4 x \Big(
\frac{1}{2} F_{\mu\nu}^2 + \alpha_0 B^2 - 2 \partial_\mu B\, A^\mu
-\nonumber\\
&&\qquad\qquad\qquad\qquad
- i \alpha_0 B \{ c \, \bar c \}
- 2i \partial_\mu {\bar c}\, {\cal D}^\mu c + \alpha_0 c^2 {\bar c}^2 -
m_0^2 ( A_\mu^2 - 2i \alpha_0 c {\bar c} )   \Big)\qquad
\end{eqnarray}

\noindent
where

\begin{eqnarray}
&& \frac{1}{e^2} = \frac{1}{e_0^2}
- \frac{1}{8\pi^2} c_2\,\frac{11}{3}\ln \frac{M}{\mu} +O(e_0^2);
\nonumber\\
&& m = m_0 + \frac{e_0^2}{8\pi^2} c_2 \ln \frac{M}{\mu}\,m_0
\Big(\frac{\alpha_0}{8} + \frac{35}{24}\Big)+O(e_0^4);\nonumber\\
&& \alpha = \alpha_0
- \frac{e_0^2}{8\pi^2} c_2 \ln \frac{M}{\mu}
\Big(\frac{\alpha_0^2}{4}+\frac{13\alpha_0}{6}\Big) + O(e_0^4),
\end{eqnarray}

\noindent
and the renormalized fields are defined by the equations

\begin{eqnarray}
&& A^R_\mu = \Bigg[1 - \displaystyle{\frac{e_0^2}{ 8 \pi^2}}
c_2\ln \frac{M}{\mu}
\frac{(\alpha_0 -3)}{4} + O(e_0^4)\Bigg] A_\mu;\nonumber\\
&& \bar c^R \cdot c^R
= \Bigg[1 - \frac{e_0^2}{ 8 \pi^2} c_2\ln \frac{M}{\mu}
\Bigg( \frac{\alpha_0}{4}- \frac{35}{12} \Bigg) + O(e_0^4)\Bigg]
\ \bar c\cdot c;\nonumber\\
&& B^R = \Bigg[1 + \frac{e_0^2}{ 8 \pi^2} c_2\ \frac{35}{12}
\ln \frac{M}{\mu} + O(e_0^4)\Bigg] B.
\end{eqnarray}

\noindent
Because renormalized action (\ref{Renormalization_Renormalized_Action})
has exactly the same structure as original action (\ref{YM_Action}), it
is invariant under BRST transformations (\ref{BRST_Transformations}) and
anti-BRST transformations (\ref{BRST_Anti_BRST_Transformations}).

%%%%%%%%%%%%%%%%%%%%%%%%%%%%%%%%%%%%%%%%%%%%%%%%%%%%%%%%%%%%%%%%%%%%%%%%%%%

\section{Conclusion.}
\hspace{\parindent}

In this paper we found the divergent part of the one-loop effective
action for the Yang-Mills theory with BRST-invariant mass term. To
perform this calculation we used a method, which was originally
proposed by t'Hooft and Veltman \cite{tHooft} and afterwards generalized
in \cite{NuclPhys} for more complicated cases. The considered theory
turns out to be renormalizable and BRST-invariant at the quantum level.
The results obtained for running mass and parameter of the gauge coincide
with the corresponding results, found in \cite{Kondo2,Gracey} by the
diagram technique up to notations. Therefore, the calculation confirms
correction of results of \cite{Kondo2,Gracey} and also correctness of
rather complicated algorithms, constructed in \cite{NuclPhys}.

%%%%%%%%%%%%%%%%%%%%%%%%%%%%%%%%%%%%%%%%%%%%%%%%%%%%%%%%%%%%%%%%%%%%%%%%%%%

\appendix

\section{Appendix.}
\hspace{\parindent}

Second variation of action (\ref{YM_Action}) up to a multiplicative
constant can be written as

\begin{equation}
D_i{}^j =
\left(
\begin{array}{ccc}
d_{AA} & d_{Ac} & d_{A\bar c}\\
d_{cA} & d_{cc} & d_{c\bar c}\\
d_{\bar c A} & d_{\bar c c} & d_{\bar c \bar c}
\end{array}
\right)
\end{equation}

\noindent
where

\begin{eqnarray}\label{Appendix_Second_Variation}
&& d_{AA} = \Big({\cal D}_\alpha^2 + m^2\Big) \eta^{\mu\nu}
- {\cal D}^\nu {\cal D}^\mu - \frac{1}{\alpha}\partial^\mu \partial^\nu
+ {\bf F}^{\mu\nu};\nonumber\\
&& d_{Ac} = \frac{i}{2} \bar {\bf c}\,\partial^\mu
- \frac{i}{2} (\partial^\mu \bar {\bf c});\nonumber\\
&& d_{A\bar c} = - \frac{i}{2} {\bf c}\,\partial^\mu
+ \frac{i}{2} (\partial^\mu {\bf c});\nonumber\\
&& d_{cA} = - \frac{i}{2} \bar {\bf c}\,\partial^\nu
- i (\partial^\nu \bar {\bf c});\nonumber\\
&& d_{cc} = \frac{\alpha}{4} \bar {\bf c}^2;\nonumber\\
&& d_{c\bar c} = i {\cal D}_\alpha \partial^\alpha
- \frac{i\alpha}{2} {\bf B} - i\alpha m^2
- \frac{\alpha}{4} \bar {\bf c} {\bf c}
- \frac{\alpha}{2} {\bf c} \bar {\bf c};\nonumber\\
&& d_{\bar c A} =
\frac{i}{2} {\bf c}\,\partial^\nu + i (\partial^\nu {\bf c});\nonumber\\
&& d_{\bar c c} = - i\partial_\alpha {\cal D}^\alpha
-\frac{i\alpha}{2} {\bf B} + i\alpha m^2
- \frac{\alpha}{4} {\bf c} \bar {\bf c}
- \frac{\alpha}{2} \bar {\bf c} {\bf c};\nonumber\\
&& d_{\bar c \bar c} = \frac{\alpha}{4} {\bf c}^2.
\end{eqnarray}

\noindent
Here we used the following notations:

\begin{equation}
{\bf B} \equiv -ie B^a T^a,\qquad (T^a)_{bc} = - i f_{abc},
\quad \mbox{etc}.
\end{equation}

Coefficients at the second derivatives in equation
(\ref{Appendix_Second_Variation}) form the matrix $K$,
coefficients at the first derivatives form the matrix $S$ and
terms without derivatives form the matrix $W$.

%%%%%%%%%%%%%%%%%%%%%%%%%%%%%%%%%%%%%%%%%%%%%%%%%%%%%%%%%%%%%%%%%%%%%%%%%%%

\end{document}